\documentclass[runningheads]{llncs}
\usepackage[T1]{fontenc}
\usepackage{graphicx}
\usepackage{amsmath}

\usepackage{cite}
\usepackage{tabularx, booktabs}
\usepackage{algorithm}
\usepackage{algorithmic}
\usepackage[switch]{lineno}
\usepackage{enumitem}
\usepackage{pifont}
\usepackage{tcolorbox}
\usepackage{multirow}

\usepackage{marvosym}

\usepackage{listings}
\newsavebox{\codeboxOne}
\newsavebox{\codeboxTwo}

\newtcolorbox{conclusionbox}[1]{
}

\begin{document}
\title{DeepCRCEval: Revisiting the Evaluation of Code Review Comment Generation}
\titlerunning{DeepCRCEval: Revisiting the Evaluation of CRCGs}
\author{
Junyi Lu\inst{1,2} \and
Xiaojia Li\inst{3} \and
Zihan Hua\inst{1,2} \and
Lei Yu\inst{1,2} \and 
Shiqi Cheng\inst{1} \and 
Li Yang\inst{1}\textsuperscript{(\Letter)} \and
Fengjun Zhang\inst{1} \and 
Chun Zuo\inst{4}
}

\authorrunning{Lu et al.}

\institute{
Institute of Software, Chinese Academy of Sciences, Beijing, China \\
\email{\{yulei2022,chengshiqi,yangli2017,fengjun\}@iscas.ac.cn}\\ 
\and
University of Chinese Academy of Sciences, Beijing, China \\
\email{\{lujunyi21,huazihan22\}@mails.ucas.ac.cn}
\and 
Kuaishou Technology, Beijing, China \\
\email{lixiaojia03@kuaishou.com}
\and
Sinosoft Company Limited, Beijing, China \\
\email{zuochun@sinosoft.com.cn}
}
\maketitle              % typeset the header of the contribution
\begin{abstract}
Code review is a vital but demanding aspect of software development, generating significant interest in automating review comments. Traditional evaluation methods for these comments, primarily based on text similarity, face two major challenges: inconsistent reliability of human-authored comments in open-source projects and the weak correlation of text similarity with objectives like enhancing code quality and detecting defects.

This study empirically analyzes benchmark comments using a novel set of criteria informed by prior research and developer interviews. We then similarly revisit the evaluation of existing methodologies. Our evaluation framework, DeepCRCEval, integrates human evaluators and Large Language Models (LLMs) for a comprehensive reassessment of current techniques based on the criteria set. Besides, we also introduce an innovative and efficient baseline, LLM-Reviewer, leveraging the few-shot learning capabilities of LLMs for a target-oriented comparison.

Our research highlights the limitations of text similarity metrics, finding that less than 10\% of benchmark comments are high quality for automation. In contrast, DeepCRCEval effectively distinguishes between high and low-quality comments, proving to be a more reliable evaluation mechanism. Incorporating LLM evaluators into DeepCRCEval significantly boosts efficiency, reducing time and cost by 88.78\% and 90.32\%, respectively. Furthermore, LLM-Reviewer demonstrates significant potential of focusing task real targets in comment generation.

\end{abstract}
\section{Introduction}
 % Since its establishment by Fagan in 1976, code review has been a cornerstone of software development, instrumental in elevating code quality and identifying defects \cite{fagan2002design}. Within both Open-Source Software (OSS) \cite{rigby2008open,rigby2014peer,rigby2013convergent} and industrial applications \cite{sadowski2018modern,shan2022using}, Modern Code Review (MCR) has evolved into a vital, streamlined process. This involves: (A) developers proposing new code through pull requests; (B) reviewers assessing these proposals; (C) providing feedback on potential issues or areas for improvement; (D) authors refining their code accordingly. A pivotal aspect of MCR is the generation of insightful, constructive comments, essential for guiding developers and pinpointing problems, which is further elaborated in Section \ref{sec: additional background}.

Since Fagan's pioneering work on software inspections in 1976 \cite{fagan2002design}, code review practices have evolved significantly. Early software inspections were formal and resource-intensive, which limited their widespread adoption in the industry. In contrast, Modern Code Review (MCR) has become a vital and streamlined process in both Open-Source Software (OSS) \cite{rigby2008open,rigby2014peer,rigby2013convergent} and industrial applications \cite{sadowski2018modern,shan2022using}. MCR typically involves developers submitting code changes for review, which reviewers then assess to provide feedback on potential issues or improvements. This process may involve mechanisms such as pull requests, changesets, or code patches, depending on the tools and platforms used. A pivotal aspect of MCR is the generation of insightful, constructive comments, essential for guiding developers and pinpointing problems, which is further elaborated in Section \ref{sec: additional background}.

Despite its effectiveness, MCR remains resource-intensive \cite{yang2016mining}. This challenge has spurred the exploration of automating code review comments to alleviate labor demands\cite{hua2024survey}. Initial attempts focused on retrieval-based methods using existing comments as references \cite{gupta2018intelligent,siow2020core}. With the progression of deep learning, the emphasis has shifted to generative approaches. A variety of Code Review Comment Generators (CRCGs) have emerged, notably initiated by Tufano et al. with a T5 transformer architecture \cite{tufano2021towards}, later augmented with code-technical language pre-training \cite{tufano2022using}. Following this trend, models like CodeReviewer \cite{li2022automating} and AUGER \cite{li2022auger} have further advanced the field, integrating specific pre-training for code review and utilizing review tags to enhance accuracy. Parallel developments include CommentFinder \cite{hong2022commentfinder}, offering an efficient retrieval-based solution, and CCT5 \cite{lin2023cct5}, which underscores the importance of considering code changes in comment generation. Later, with the emergence of LLMs, Lu et al. \cite{lu2024exploring} investigate the factors influencing the code review process for both traditional pre-trained language models and LLMs, while Llama-Reviewer \cite{lu2023llama} marks the first attempt at training code review tasks on LLMs in a parameter-efficient approach. Nevertheless, these models typically employ text similarity metrics such as BLEU\cite{papineni2002bleu} and ROUGE\cite{lin2004rouge} for evaluation.

We question the suitability of text similarity as the primary metric for assessing code review comment automation. The dependability of human-written comments in OSS, commonly used as benchmarks, is often subject to scrutiny \cite{bacchelli2013expectations,bosu2015characteristics,bosu2017process,kononenko2016code,rahman2017predicting,hasan2021using,yang2023evacrc}. These comments can be arbitrary and inconsistent, at times offering little more than basic queries or directives like ``Why do we need this?'' or ``Remove this.'' Unlike traditional text-to-text tasks such as summarization or translation, which seek semantic equivalence, code review comments aim to aid in defect detection and code refinement, requiring deeper insight \cite{bacchelli2013expectations}. Hence, text similarity, as an indirect measurement, falls short of capturing the essence.

\noindent\textbf{Analysis of Benchmark Comments.}\indent The reliability of text similarity metrics for evaluating review comments hinges on the accuracy of the reference text. Yet, the quality and validity of benchmark comments in major datasets, such as the CodeReviewer (CRer) \cite{li2022automating} and Tufano datasets \cite{tufano2022using}, often remain ambiguous. To address these ambiguities and provide a more nuanced understanding, we conducted an empirical study analyzing these datasets from four dimensions: 1) Quality: Assessing comments against established quality standards; 2) Category: Evaluating the effectiveness of comments in identifying defects and suggesting improvements, or other potential roles; 3) Tone: Analyzing whether comments clearly state issues or are merely interrogative; 4) Context: Determining if the comments are sufficiently supported by the associated code. For Quality evaluation, we developed criteria based on existing literature about developers' views on code review quality \cite{kononenko2016code}, supplemented by our semi-structured interviews and card sorting exercises using affinity diagrams.

\noindent\textbf{DeepCRCEval}\indent Stemming from our analysis of benchmark comments, we developed DeepCRCEval, an innovative evaluation framework incorporating both human and LLM evaluators. This framework utilized the criteria of high quality comments identified in the former analysis, and use both scoring and ranking to compare performances. It is geared towards identifying the intrinsic merits of review comments, moving away from the indirect approach of text similarity. 

\noindent\textbf{LLM-Reviewer}\indent To validate the effectiveness of direct metrics in line with code review objectives, we propose LLM-Reviewer, a lightweight, training-free baseline tool. Traditional methods, reliant on text similarity metrics, may not capture the true essence of code review. In contrast, by harnessing the potential of Large Language Models (LLMs), LLM-Reviewer employs few-shot learning with meticulously crafted prompts. Unlike previous methods, LLM-Reviewer directly addresses the actual aims of code review and controlled the comments' quality with criterion guidance in prompt, more closely mirroring developers' concept of effective review commentary. This baseline also serves as a benchmark to gauge the effectiveness of evaluation frameworks in differentiating high and low-quality review comments.

\noindent\textbf{Revisiting the Evaluation of CRCGs}\indent Utilizing DeepCRCEval as our evaluation framework and LLM-Reviewer as our baseline, we reassessed the performance of current state-of-the-art (SOTA) CRCGs. Given the training-free nature of LLM-Reviewer, it was anticipated that other tuned methods would perform at least comparably.  However, our findings, derived from both human and LLM evaluators, reveal a considerable shortfall in the performance of existing SOTA CRCGs compared to this benchmark. Our analysis of 1,000 typical code snippets containing defects or code smells demonstrated LLM-Reviewer's proficiency in consistently pinpointing issues, providing well-rounded comments encompassing problem identification, detailed explanation, and possible solutions. In contrast, SOTA CRCGs often produced nonspecific and ambiguous comments, falling short of the review's goals.

\noindent\textbf{Contributions}\indent Our significant contributions include:
\begin{enumerate}
\item Unveiling the biases in evaluating state-of-the-art (SOTA) CRCGs due to reliance on inappropriate metrics.
\item Developing DeepCRCEval, a versatile evaluation framework employing either human or LLM evaluators, which concentrates on the core essence of comments rather than indirect text similarity measures.
\item Introducing LLM-Reviewer, a pioneering, training-free baseline for code review comment automation.
\item Empirically demonstrating that existing SOTA CRCGs are outperformed by the training-free LLM-Reviewer, indicating potential for improvement.
\item Materials publicly available at https://zenodo.org/records/10511726.
\end{enumerate}

The rest of the paper is structured as follows: Section \ref{sec: background} presents the background of our research. Section \ref{sec: overview} introduces the overview of the study and our research questions. Section \ref{sec: analysis} details our analysis of benchmark comments, which aims to prove the limitations of current text similarity measurements. Section \ref{sec: revisiting} illustrates our approach to revisiting the evaluation of code review comment generators, including our dual-granularity human/LLM evaluation framework DeepCRCEval, and a prompt-based LLM baseline LLM-Reviewer. Section \ref{sec: findings} shows the results of the reevaluation, as well as key findings and discussions. Section \ref{sec: related work} introduces related work on the evaluation of code review comments and the differences between these and our work. Section \ref{sec: conclusion} concludes the paper.

\section{Background} \label{sec: background}

\subsection{CRCGs and Their Evaluation} \label{sec: CRCGs}
\begin{figure}[htbp]
    \centering
    \includegraphics[width=0.8\linewidth]{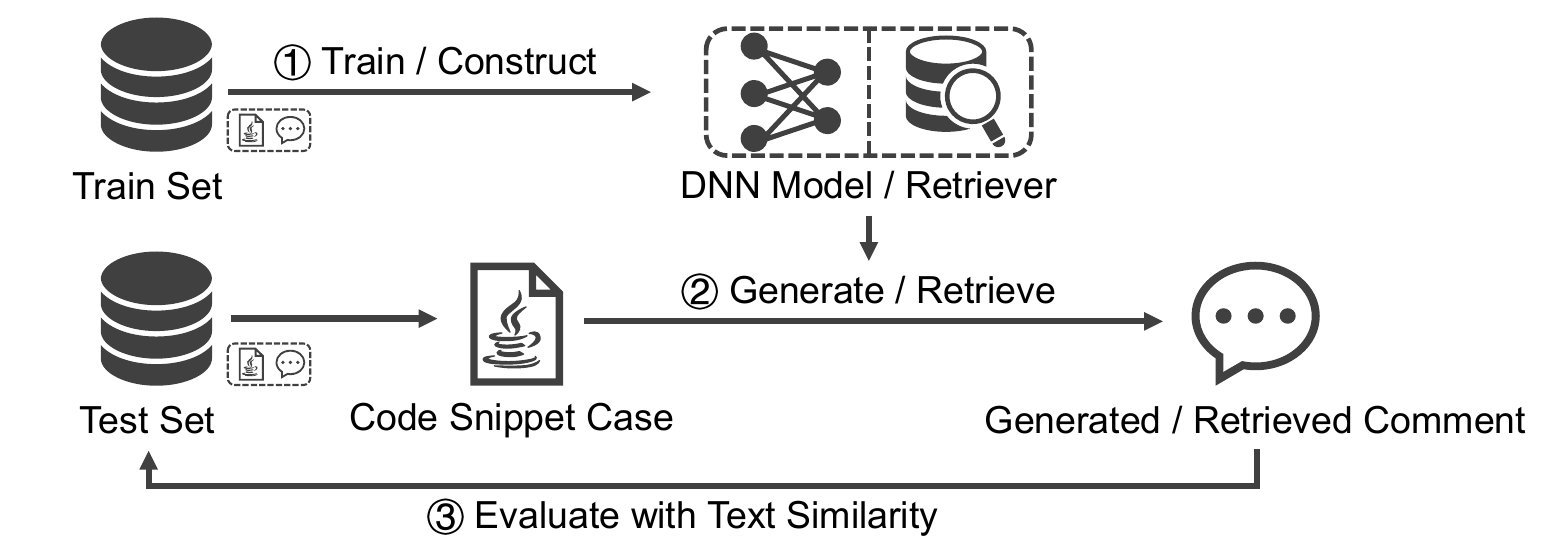}
    \caption{The overall workflow of learning a deep neural network (DNN) model or retriever to automate code review.}
    \label{fig:overall_workflow_previous}
\end{figure}

For code review automation, researchers predominantly utilize deep learning or information retrieval techniques as code review comment generators (CRCGs) to automatically generate comments for given code snippets. Despite some variations, Figure \ref{fig:overall_workflow_previous} depicts the typical workflow for training a deep neural network (DNN) model or constructing a retriever for automating code review. Initially, \ding{192} a DNN model or a retriever is trained or established using a dataset of code-comment pairs from OSS projects, learning the semantic relationships between code and comments. Then, \ding{193} for each test case code snippet, the model or retriever generates or retrieves a comment employing specific decoding or retrieval techniques. Finally, \ding{194} the produced comments are compared against the original ones from the test set, also sourced from OSS projects. Common comparison metrics include BLEU\cite{papineni2002bleu} and ROUGE\cite{lin2004rouge}, where BLEU assesses the precision of machine translations by measuring the match of N-grams, and ROUGE evaluates the recall of machine-generated summaries based on N-gram co-occurrence. These metrics also guide the training of DNNs.

However, the quality and validity of the original comments extracted from OSS projects are questionable. For example, the comments in Table \ref{tab:example_comments} are not suitable for use as code review comments by \textbf{models}, although they might be meaningful for human reviewers. Specifically, ``Why do we need this?'' is indeed meaningful if it opens a dialogue when by humans. However, CRCGs generate comments only once, without a dialogue-like interaction. If a model outputs this comment, the code author cannot acquire further knowledge as the process has concluded. ``Remove this line'' provides specific action suggestions. However, without reasons, it could confuse code authors. If by humans, this could be clarified by asking further questions, but for CRCGs the process is already complete. Besides, these two comments are both too general. They lack context and explanation, making it applicable to any code.

Moreover, the text similarity used for comparison are indirect measures of the primary objective, which is to find defects or enhance code. Unlike tasks such as text translation where similar input leads to similar output, review comments can be arbitrary. A same issue can be represented differently, while similar representations might have totally different meanings. Therefore, we posit that there might be an overestimation of the effectiveness of current SOTA CRCGs.

\begin{table}[ht]
    \centering
    \scriptsize
    \caption{Example of unsuitable comments for machine code review.}
    \label{tab:example_comments}
    \begin{tabular}{p{12cm}}
         \specialrule{0.1em}{0em}{0em} 
         It's a race to see who merges first, because I bet one of my PRs will cause a conflict with this \\ \hline
         why do we need this \\ \hline
         remove this line \\ \hline
         While I would have liked to resolve all of the N+1 issues related to profiles in this PR this has proved more difficult than anticipated and there's a lot of other good stuff in this PR that I'd really like to get out, so we can keep this around a bit longer. \\ \hline
         Can we avoid change names of variable for now? This change is not purely style change and might make those who maintains private patches harder. Such change can be done separately with a more fine grained approach. \\ \hline
         I seem to remember we spoke on this earlier in the opposite sense, saying that we want the shortcuts to be added even if they are not displayed. \\ \hline
         The same code has been used a few times throughout this file. \\ \specialrule{0.1em}{0em}{0em} 
    \end{tabular}
\end{table}

\subsection{Large Language Models for Evaluations}

Large language models have shown capabilities similar to those of human evaluators\cite{li2024generation}. Studies have demonstrated that large language models like GPT-4 can achieve higher agreement than human evaluators \cite{zheng2023judging, dubois2023alpacafarm, alpaca_eval, Aviary} for ranking preference and scoring. For instance, in the evaluation of general text generation, GPT-4's agreement rate with human evaluations (85\%) surpasses the rate of inter-human agreement (81\%) \cite{zheng2023judging}, a finding echoed in another dataset comparison (69.2\% vs. 65.7\%) \cite{alpaca_eval}. Considering this, our evaluation framework is based on these works but is more granular and lightweight. We use a chain-of-thought template to inject task-specific knowledge for better scoring and ranking.

Since there is no prior proof of the effectiveness of LLM evaluators in the task of code review comment generation, we adopt both human and LLM evaluators for all evaluation parts of our paper. The LLM evaluators serve as an auxiliary certification, expanding the scope of verification.

\section{Overview and Research Questions} \label{sec: overview}

\begin{figure*}[htbp]
    \centering
    \includegraphics[width=0.9\linewidth]{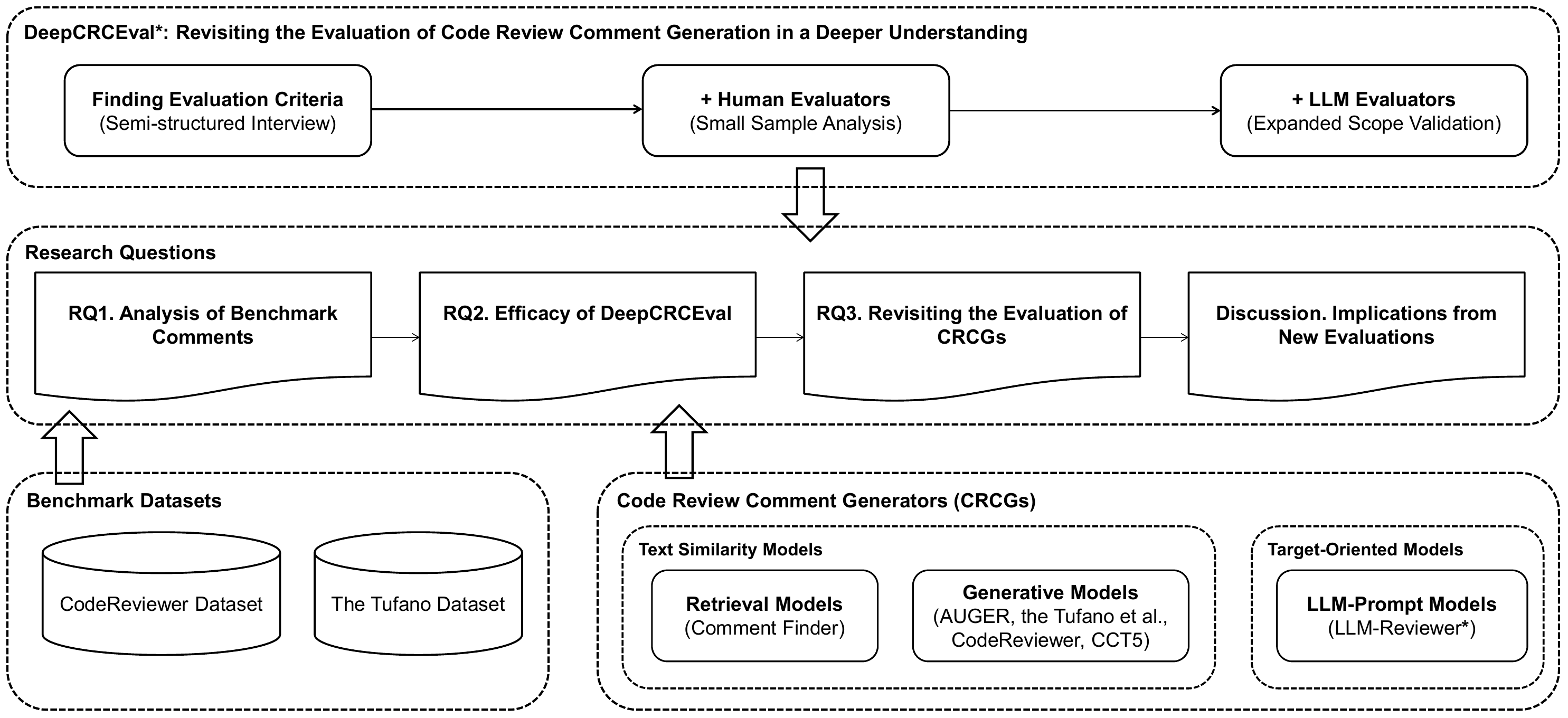}
    \caption{Overview of our study. * indicates frameworks or models we newly proposed.}
    \label{fig:overview_of_study}
\end{figure*}

Automated code review has garnered significant interest from researchers, yet its evaluation effectiveness has not been sufficiently addressed. This oversight leads to a disconnect between the task objectives and the training processes used. This paper aims to highlight the importance of effective evaluation in code review automation. As depicted in Figure \ref{fig:overview_of_study}, we developed \ding{172} DeepCRCEval, a new evaluation framework incorporating both human and LLM evaluators, and \ding{173} LLM-Reviewer, a lightweight, training-free, and target-oriented baseline tool. Our study focuses on: \textbf{1)} verifying the shortcomings of current evaluation metrics, \textbf{2)} manually reassessing the evaluation of existing code review comment generators, \textbf{3)} integrating LLM-based alternative evaluators for an extended scope of validation, and \textbf{4)} identifying potential improvement directions stemming from the misalignment of task objectives and training processes. The latter two RQs aim to set directions for future research. We next introduce the research questions we aim to investigate and their relationships.

\textbf{RQ1. Analysis of Benchmark Comments: Are the foundations of current evaluation metrics reliable?} The reliability of text similarity metrics for evaluating review comments depends on the quality and validation of the reference texts. We first analyze the benchmark reference review comments from four perspectives: quality, category, tone, and context. We present results for each aspect and summarize the relationships between these aspects.

\textbf{RQ2. Efficacy of DeepCRCEval: Why do we claim that DeepCRCEval provides a deeper evaluation, and why do we integrate LLM evaluators?} Before presenting the new results obtained in RQ3, we first examine the differences between our DeepCRCEval and traditional text similarity metrics. Additionally, we evaluate the advantages and disadvantages of using LLM evaluators for expanded scope validation. We report on the efficiency of LLM evaluators and their agreement with human evaluators.

\textbf{RQ3. Revisiting the Evaluation of CRCGs: What are the actual performances of current CRCGs beyond simple text similarity metrics?} By challenging the current text similarity metrics in RQ1, we aim to investigate the actual performances of current CRCGs more deeply. Using our newly proposed evaluation framework, DeepCRCEval, and the new target-oriented model, LLM-Reviewer, as a baseline, we increase the distinction in our analyses. We analyze results from human evaluators as a small sample analysis and from LLM evaluators for expanded scope validation.

\textbf{Discussion. Implications from New Evaluations: What can we learn to guide further research?} The new evaluations aim to guide future research in code review comment generation and propose potential improvement directions. We present the implications derived from our study with case studies.

\section{RQ1. Analysis of Benchmark Comments} \label{sec: analysis}

% Our investigation challenges the quality and validity of benchmark comments used as reference standards. We aim to determine the extent to which these comments may be unsuitable for reference purposes.

\subsection{Aspects for Analysis}
To thoroughly explore previous code review comment datasets, we defining aspects with a qualitative and quantitative process \cite{creswell2017research}, which draws on varied data sources for comprehensive insights, including the review of previous study \cite{kononenko2016code}, our semi-structured interviews \cite{lindlof2017qualitative} with seven industry developers, and subsequent card sort and affinity diagram by the authors of this paper. The detailed process are shown in Section \ref{sec: finding quality evaluation criteria}. These gained criteria are utilized for assessing benchmark comment quality, which will also for the subsequent reevaluation. 

% In addition, this section extends the analysis by incorporating perspectives related to code content and user experience, including the categories, tone, and context, to provide a more comprehensive evaluation.

\subsubsection{Quality} \label{sec: quality}

The multi-step study introduced in Section \ref{sec: finding quality evaluation criteria} culminated in identifying nine key criteria for evaluating the effectiveness of code review comments:
\begin{enumerate}[label=\textbf{C\arabic*. }, labelwidth=*, labelsep=0em, leftmargin=!, align=left, start=1]
    \item \textbf{Readability:} Clear, easily understandable language.
    \item \textbf{Relevance:} Directly related to the specific code.
    \item \textbf{Explanation Clarity:} Clear elucidation of the issues identified.
    \item \textbf{Problem Identification:} Accurate pinpointing and articulation of bugs.
    \item \textbf{Actionability:} Practical advice for addressing identified issues.
    \item \textbf{Completeness:} Coverage of all issues in the code for comprehensive review.
    \item \textbf{Specificity:} Focus on specific code issues, avoiding generic statements.
    \item \textbf{Contextual Adequacy:} Comments pointing out exact issue locations.
    \item \textbf{Brevity:} Conciseness, conveying essential information without verbosity.
\end{enumerate}

\subsubsection{Category}

We evaluated the comments' ability to detect defects or suggest code improvements using a classification system. This involved adopting the nine categories proposed by Bacchelli et al. \cite{bacchelli2013expectations}, such as Code Improvement, Understanding, and Social Communication, and introducing an additional category, ``Meaningless Text'', for extremely low-quality, uninformative comments.

\subsubsection{Tone and Context}

To examine factors influencing comment validation, we conducted a manual inspection using the Nominal Group Technique (NGT), involving three authors of this paper. Each participant initially prepared individual opinions, later discussed in a structured meeting. The focus was on the aspects with the highest consensus:
\begin{itemize}
    \item \textbf{\ding{172} Tone:} Interrogative comments were noted to be less effective, as they often raise questions rather than providing specific, formalized feedback. For example, a comment like ``Why this?'' is less valuable for identifying defects or suggesting improvements.
    \item \textbf{\ding{173} Context:} Comments requiring understanding beyond the provided code snippet (e.g., at the file level) were found challenging, especially in the context of automated code review techniques and their datasets. Comments such as ``used a few times throughout this file'' lack clarity when only a single method is given as input.
\end{itemize}

\subsection{Analysis Methodology}

Analyzing code review comments is a nuanced and labor-intensive task. To manage this, we sampled 100 comments from each of the two primary datasets in this domain: the Tufano dataset \cite{tufano2022using} and the CodeReviewer dataset \cite{li2022automating}, abbreviated as Tufano and CRer, respectively. According to the average reliability of 93\% for humans in Table 6, the margin of error for 95\% confidence level of 100 sample size is within 5\%. The Tufano dataset is a monolingual, function-level Java dataset, while the CRer dataset is multilingual and at the diff granularity. Both are constructed from large-scale open-source software repositories and widely utilized in numerous studies. The quality and category is finished with a human scoring system created using QT and and a Delphi Method variant\cite{dalkey1963experimental} by five master's and doctoral students, respectively. The Tone and context was conducted using the aforementioned NGT sessions by three authors of this study. The detailed process are shown in Section \ref{sec: Detailed Analysis Methodology for Dateset Comment Quality}.

\subsection{Results of Analysis} \label{sec: results of analysis}

We first present the results from each aspect, and then summarize the relationships between each aspect using a Venn diagram in Section \ref{sec: summary of analysis}.

\subsubsection{Quality} \label{sec: quality results}

Our analysis, illustrated in the upper part of Table \ref{tab:quality}, indicates that while OSS review comments typically exhibit good readability and brevity, they frequently lack in other critical aspects, reflecting issues of arbitrariness, incompleteness, and irregularity. We designated scores \textbf{below 6 (out of 10)} as poor performance indicators for each aspect. The lower part of Table \ref{tab:quality} shows the proportion of comments scoring poorly in each aspect. This data highlights deficiencies of comments in aspects other than readability, completeness, and brevity. For example, low scores in explanation clarity or actionability hint at inadequate detail or absence of constructive suggestions, while low relevance, contextual adequacy, or specificity scores suggest a tendency towards vagueness or irrelevance to the code context. LLM evaluations were also conducted, similar to Section \ref{sec: DeepCRCEval}, with detailed results available in Table \ref{tab:quality_by_llms}.

\begin{table}[ht]
\centering
\scriptsize
\caption{Average quality of comments (\ding{182} upper part, 1-10) and percentage of low-quality cases (\ding{183} lower part, 0\%-100\%) in OSS datasets. C1-C9 represent aspects mentioned in Section \ref{sec: quality}.}
\label{tab:quality}
\begin{tabular}{@{}llrrrrrrrrr@{}}
\toprule
                    & Dataset & C1   & {\color[HTML]{FF0000}C2}   & {\color[HTML]{FF0000} {C3}}   & {\color[HTML]{FF0000} {C4}}   & {\color[HTML]{FF0000} {C5}}   & C6   & {\color[HTML]{FF0000} {C7}}   & {\color[HTML]{FF0000} {C8}}   & C9   \\ \midrule
                    & Tufano  & 8.68 & {\color[HTML]{FF0000}7.07} & {\color[HTML]{FF0000} {4.70}} & {\color[HTML]{FF0000} {5.40}} & {\color[HTML]{FF0000} {5.81}} & 7.61 & {\color[HTML]{FF0000} {6.43}} & {\color[HTML]{FF0000} {4.89}} & 8.99 \\
\multirow{-2}{*}{\ding{182}} & Crer    & 9.21 & {\color[HTML]{FF0000}6.41} & {\color[HTML]{FF0000} {5.23}} & {\color[HTML]{FF0000} {5.78}} & {\color[HTML]{FF0000} {5.08}} & 7.40 & {\color[HTML]{FF0000} {6.36}} & {\color[HTML]{FF0000} {6.32}} & 9.03 \\ \midrule
                    & Tufano  & 7\%  & {\color[HTML]{FF0000}31\%} & {\color[HTML]{FF0000} {64\%}} & {\color[HTML]{FF0000} {55\%}} & {\color[HTML]{FF0000} {46\%}} & 20\% & {\color[HTML]{FF0000} {34\%}} & {\color[HTML]{FF0000} {63\%}} & 4\%  \\
\multirow{-2}{*}{\ding{183}} & Crer   & 1\%  & {\color[HTML]{FF0000}48\%} & {\color[HTML]{FF0000} {57\%}} & {\color[HTML]{FF0000} {52\%}} & {\color[HTML]{FF0000} {64\%}} & 31\% & {\color[HTML]{FF0000} {35\%}} & {\color[HTML]{FF0000} {43\%}} & 0\%   \\ \bottomrule
\end{tabular}
% \vspace{-0.3cm}
\end{table}

\subsubsection{Category}

The classification of comment categories in OSS projects, summarized in Table \ref{tab:distribution_of_categories}, reveals distinct distributions. In the Tufano dataset, code improvements and defects constitute 64\%, while in the CRer dataset, they comprise only 39\%. This distribution aligns with previous findings on the proportion of practically useful comments \cite{bosu2015characteristics}. Comments outside these categories may hold value in specific human code review scenarios but do not align with the core objectives of machine code reviews. For example, comments on deferring tasks to future pull requests, unrelated to the current code, offer limited value.

\begin{table}[ht]
\centering
\scriptsize
\caption{Distribution of comment categories in OSS datasets.}
\label{tab:distribution_of_categories}
\begin{tabular}{@{}lrr@{}}
\toprule
                                        & the Tufano dataset          & the CRer dataset            \\ \midrule
{\color[HTML]{0070C0} Code Improvement} & {\color[HTML]{0070C0} 43\%} & {\color[HTML]{0070C0} 26\%} \\
Understanding                           & 5\%                         & 19\%                        \\
Social Communication                    & 12\%                        & 18\%                        \\
{\color[HTML]{0070C0} Defects}          & {\color[HTML]{0070C0} 21\%} & {\color[HTML]{0070C0} 13\%} \\
External Impact                         & 2\%                         & 1\%                         \\
Testing                                 & 2\%                         & 3\%                         \\
Review Tool                             & 2\%                         & 1\%                         \\
Knowledge Transfer                      & 1\%                         & 8\%                         \\
Misc                                    & 1\%                         & 3\%                         \\
Meaningless Text                        & 11\%                        & 8\%                         \\ \bottomrule
\end{tabular}
% \vspace{-1cm}
\end{table}

\subsubsection{Tone and Context} \label{sec: tone and context results}

The analysis shows a significant presence of interrogative comments—38\% in the Tufano dataset and 46\% in the CRer dataset. Besides, a substantial portion of comments—45\% in the Tufano dataset and 54\% in the CRer dataset—required out-of-method or out-of-hunk context.

\subsubsection{Summary of Analysis} \label{sec: summary of analysis}

\begin{figure}[t]
    \centering
    \includegraphics[width=\linewidth]{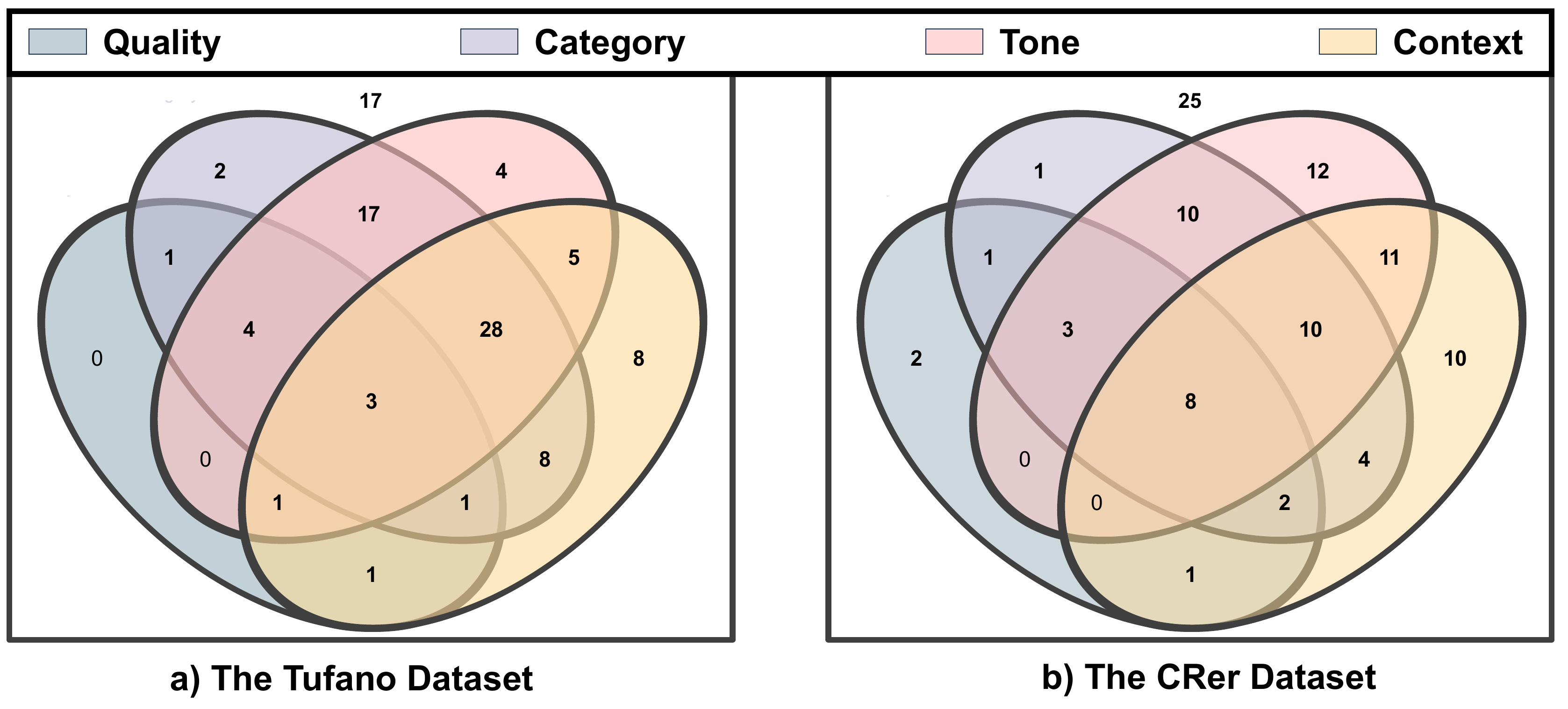}
    \caption{4-group Venn diagrams showing the overlap of suitable quality, category, tone, and context in comments.}
    \label{fig:venn_diagram}
    % \vspace{-0.4cm}
\end{figure}

The Venn diagrams in Figure \ref{fig:venn_diagram} for each dataset summarize the interplay of various factors impacting the suitability of reference comments. A notable finding is that only a small fraction of comments in these datasets—3\% in the Tufano dataset and 8\% in the CRer dataset—qualify as ideal references. Furthermore, even within this subset, many comments only address minor issues like typos or simple syntax errors.

\begin{tcolorbox}
\textbf{Response to RQ1:} The uncertain quality and validity of benchmark comments undermine their suitability as reference standards, casting doubt on the effectiveness of text similarity as a metric.
\end{tcolorbox}
  
\section{Revisiting the Evaluation of CRCGs} \label{sec: revisiting}

In light of the inadequacy of text similarity metrics for evaluating Code Review Comment Generators (CRCGs), this section introduces our novel evaluation framework, DeepCRCEval, and a target-oriented baseline, LLM-Reviewer, to reassess the performance of CRCGs.

\subsection{DeepCRCEval} \label{sec: DeepCRCEval}

Drawing from the criteria established in Section \ref{sec: quality}, we developed DeepCRCEval, an evaluation framework designed to rigorously analyze the quality of generated comments. DeepCRCEval employs both human and Large Language Model (LLM) evaluators. Initially, we conducted human evaluations on a sampled test set due to cost considerations, then extended the evaluation to the entire test set using LLMs.

\subsubsection{Human Evaluators}

For human evaluations, we developed an executable file using QT, similar to the one in Section \ref{sec: results of analysis} but enhanced with a comparative ranking task. This tool is also available in our open-source repository.

\subsubsection{LLM Evaluators} \label{sec: llm evaluators}

Recognizing the limitations of human scoring, such as cost and time intensity, and challenges in assessing certain aspects like completeness, we incorporated LLM evaluators. Our methodology utilizes a thought-chain-enhanced prompt template, adapted from recent research \cite{zheng2023judging}:

\begin{itemize}
    \item \textbf{Domain-Based Scoring:} In contrast to earlier methods that used a single overall score, we introduced domain-specific scoring across nine different dimensions, allowing for a more detailed and nuanced assessment.
    \item \textbf{Efficient Sorting:} We optimized resource usage by streamlining pairwise comparisons into a consolidated overall sorting, simplifying the process.
    \item \textbf{Chain-of-Thought Integration:} Chain-of-thought (CoT) module was embedded to link scoring and ranking tasks coherently, enhancing the logical flow of evaluations.
\end{itemize}

The structure of our enhanced prompt template is:
\begin{align}
    P_{Eval} &= Des_{Scoring} + G + Obj + Des_{Ranking} + F \\
    F &= F_{Scoring} + CoT + F_{Ranking}
\end{align}
where \( Des_{Scoring} \) and \( Des_{Ranking} \) denote the task descriptions for scoring and ranking, respectively. \( G \) comprises guidelines, including notes and criteria descriptions, while \( Obj \) represents the objects being evaluated. \( F \) outlines the format for the generation, integrating a CoT section for enriched explanations. We delineate each component with ``\#\#\#'' for clarity. The detailed prompt template is shown in Table \ref{tab:prompt_template_eval} in the Appendix. To minimize bias, evaluations were performed twice for each case, once in descending and once in ascending order.

% \begin{figure}
%     \centering
%     \includegraphics[width=\linewidth]{images/prompt_template_evaluator.pdf}
%     \caption{Detailed prompt template for LLM evaluators.}
%     \label{fig:prompt_template_eval}
% \end{figure}

\subsection{Selected CRCGs}

For our analysis, we first selected current state-of-the-art (SOTA) CRCGs published in top-tier conferences. This includes models like Tufano et al. \cite{tufano2022using}, AUGER \cite{li2022auger}, CodeReviewer \cite{li2022automating}, and CCT5 \cite{lin2023cct5}, primarily based on DNN models, alongside CommentFinder \cite{hong2022commentfinder}, which uses retrieval techniques. Notably, all these CRCGs have been trained or constructed using data from OSS projects and guided by text similarity metrics, which may not align well with the primary goal of identifying defects or enhancing code.

To evaluate CRCG performance effectively, an appropriate baseline is crucial. We sought a baseline that directly targets defect detection and code improvement. Given the absence of such a baseline in existing literature, we introduce LLM-Reviewer, a novel, straightforward, and fair baseline. It resembles DNNs but does not require additional training. 

\subsection{LLM-Reviewer}

\begin{figure}[htbp]
    \centering
    \includegraphics[width=0.8\linewidth]{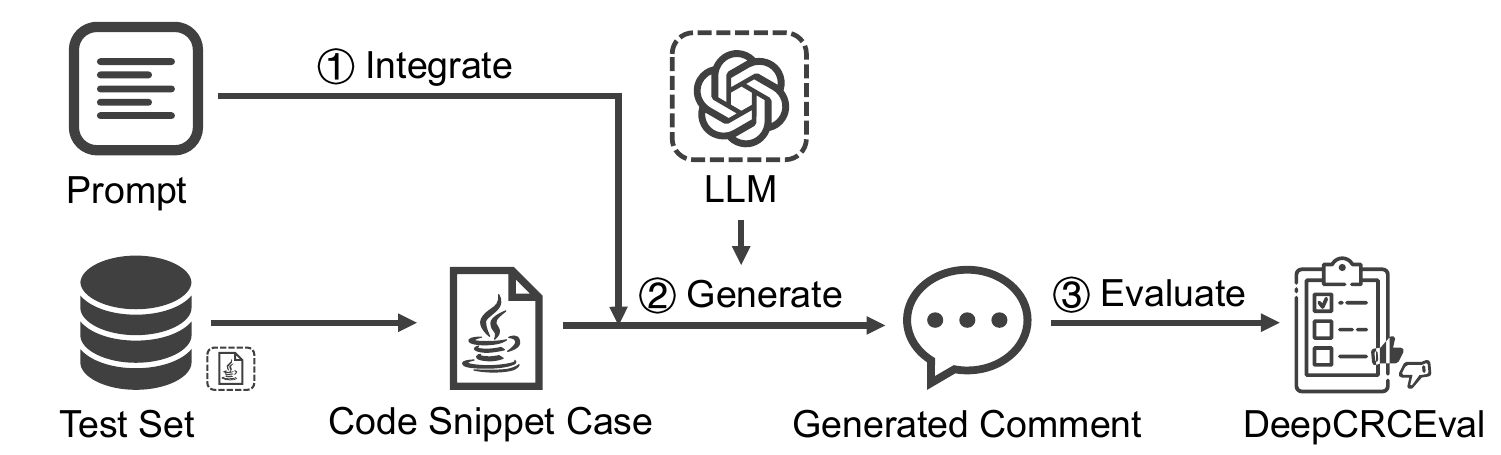}
    \caption{The overall workflow of LLM-Reviewer.}
    \label{fig:llm_reviewer}
    % \vspace{-0.3cm}
\end{figure}

Figure \ref{fig:llm_reviewer} illustrates the workflow of LLM-Reviewer. Distinct from traditional CRCGs, LLM-Reviewer operates without the need for a training set. The process for each new code snippet is: \ding{192} Integration of the code with a pre-defined prompt. \ding{193} Feeding this combined input into the selected LLM to generate a comment. \ding{194} Direct evaluation of the generated comment's quality using DeepCRCEval.

The pre-defined prompt is bolstered with specific guidance and few-shot instructions. The utilized in-context prompt is:
\begin{align}
    P_{Reviewer} &= Des_{Gen} + G_{Gen} + Demo^{k}
\end{align}
where \( Des_{Gen} \) offers concise directives for generating code review comments, \( G_{Gen} \) encompasses guidelines including notes and criteria descriptions (as discussed in Section \ref{sec: quality}), and \( Demo^{k} \) features k randomly chosen demonstrations, adhering to k-shot learning principles. For this study, we opted for \( k=3 \), balancing input length with informative content. Each prompt component is clearly separated by ``\#\#\#'' as detailed in Table \ref{tab:prompt_template_llm_reviewer}.

% \begin{figure}[ht]
%     \centering
%     \includegraphics[width=\linewidth]{images/prompt_template_reviewer.pdf}
%     \caption{Detailed prompt template for LLM-Reviewer.}
%     \label{fig:prompt_template_llm_reviewer}
% \end{figure}

This prompt is then integrated with the code snippet under review before being fed into the LLM for comment generation. It is worth noting that we use a similar prompt template to the one used for prompt evaluation. This similarity is why we refer to LLM-Reviewer as a target-oriented model. Unlike other text similarity models, the similar prompt template constrains LLM-Reviewer's response to the expected target. Additionally, as a countermeasure against bias, we use human evaluators throughout the entire study.

\subsection{Other Experimental Details}

\subsubsection{Testing Set}

Since LLM-Reviewer does not require a training dataset, we selected a set of 1,000 code cases with typical issues for testing. These cases are processed by humans to enhance simplicity, and thus to reduce the risk of data leakage. Each case is guaranteed to contain at least one significant issue within one or a few functions, compatible with the input format of earlier CRCGs. We used ROUGE-L to remove duplicates. This test set is also accessible in our open-source repository. For baselines, we utilized their respective training sets.

\subsubsection{Implementation of Baselines}

The implementation of baseline CRCGs involved different approaches. For Tufano et al. \cite{tufano2022using}, AUGER \cite{li2022auger}, and CCT5 \cite{lin2023cct5}, we leveraged their publicly available models. CommentFinder \cite{hong2022commentfinder} was replicated using the original dataset and code, while CodeReviewer \cite{li2022automating} was obtained through fine-tuning their pre-trained model with provided scripts. All GPT-4 usages within the DeepCRCEval framework and our LLM-Reviewer were implemented using the OpenAI Python library, with a temperature setting of 0.1 and a token limit of 8192.

\section{RQ2-RQ3: Empirical Findings} \label{sec: findings}

This section presents our empirical findings, detailing the performance of various CRCGs as evaluated by our framework DeepCRCEval.

\subsection{RQ2. Efficacy of DeepCRCEval}
This subsection examines the efficacy of our DeepCRCEval framework, specifically assessing the criteria used for evaluation and the integration of LLMs.

\subsubsection{Effectiveness of Criteria}
Our evaluation framework, DeepCRCEval, surpasses traditional text similarity metrics in two key areas: discrimination and comprehensiveness.

\paragraph{Discrimination} 
Prior studies often reported marginal improvements in text similarity metrics, like a less than 1\% increase in BLEU scores. Such negligible enhancements do not reliably indicate an improvement in comment quality, as corroborated by our reevaluations. While newer baselines like CCT5 and AUGER reported improvements in text similarity, they did not surpass their predecessor, Tufano et al., in effectiveness—a conclusion also supported by our qualitative case studies. In contrast, DeepCRCEval, with its well-defined criteria, offers a higher degree of discrimination across various aspects. Moreover, our ranking process provides a direct and comparative analysis of comment quality.

\paragraph{Comprehensiveness} 
DeepCRCEval's second advantage is its ability to offer a holistic evaluation. Unlike previous studies that could not elucidate why their methods were superior using text similarity metrics, DeepCRCEval, by incorporating domain-specific criteria, sheds light on the strengths and weaknesses of different models, providing a more rounded assessment.

\subsubsection{Efficacy of LLM Evaluators}
Integrating LLMs as evaluators aims to enhance automation and minimize cost.

\paragraph{Efficiency} 
To demonstrate the efficiency of LLM evaluators compared to human evaluators, we analyzed the average time and cost per case. Human evaluators were paid \$10 per hour, while LLM evaluations were costed based on API charges (\$0.03 per 1000 input tokens, \$0.06 per 1000 output tokens). Table \ref{tab:comparison_evaluator} contrasts the average time and cost for both single comment evaluations and comparative performance analyses. The results underscore the significant reduction in time and cost achieved with LLM evaluators.

\begin{table}[t]
\centering
\tiny
\caption{Comparison of average time and cost per case per evaluator between human and LLM evaluators.}
\label{tab:comparison_evaluator}
\begin{tabular}{@{}lrrrr@{}}
\toprule
\multirow{2}{*}{Evaluator} & \multicolumn{2}{c}{Single Comment} & \multicolumn{2}{c}{Performance Comparison} \\ \cmidrule(lr){2-3} \cmidrule(l){4-5}
 & Time (s) & Cost (\$) & Time (s) & Cost (\$) \\ \midrule
Humans & 224.45 & 0.62 & 752.65 & 2.09 \\
LLMs & 25.18 & 0.06 & 68.69 & 0.17 \\
\bottomrule
\end{tabular}
\end{table}

\paragraph{Effectiveness}
We utilized the Inter-class Correlation Coefficient (ICC) to gauge the concordance between LLM and human evaluators, setting the average human scores as the reference. Table \ref{tab:agreement} details their agreements across various evaluation aspects. Specifically, for each quality metric, two coefficients are reported: one represents the agreement between human evaluators and the reference scores (Human vs. Reference), and the other captures the agreement between LLM outputs and the reference scores (LLM vs. Reference). Despite a marginally lower agreement in some areas, the LLM evaluators achieved high concordance (above 0.75) in aspects like Explanation Clarity, Problem Identification, Actionability, Specificity, and Contextual Adequacy. In aspects where LLM and human evaluators diverged, we observed distinct scoring tendencies. For example, humans were prone to extreme ratings in Readability and Relevance, while LLMs offered more evenly distributed scores. LLMs also considered grammatical and syntactic factors in Brevity, unlike humans who focused on content length and relevance.

\begin{table}[t]
\centering
\footnotesize
\caption{Levels of agreement between humans and LLMs.}
\label{tab:agreement}
\begin{tabular}{@{}lrrrrrrrrr@{}}
\toprule
Evaluators & C1 & C2 & C3 & C4 & C5 & C6 & C7 & C8 & C9 \\ \midrule
Humans & 0.89 & 0.94 & 0.94 & 0.94 & 0.95 & 0.91 & 0.94 & 0.95 & 0.90 \\
LLMs & 0.62 & 0.68 & 0.83 & 0.80 & 0.81 & 0.72 & 0.79 & 0.78 & 0.62 \\
\bottomrule\end{tabular}
\end{table}

\begin{tcolorbox}
    \textbf{Response to RQ2:} DeepCRCEval enhances the discrimination and comprehensiveness of evaluations. Employing LLMs as evaluators significantly reduces costs while maintaining a commendable level of reliability.
\end{tcolorbox}

\subsection{RQ3. Results for Baselines by DeepCRCEval}

Utilizing DeepCRCEval, as delineated in Section \ref{sec: DeepCRCEval}, we reassessed the quality of comments generated by different CRCGs. The scoring and ranking outcomes from both evaluator types were averaged to derive final results.

\subsubsection{Scoring}

The scoring results, presented in the first part of Table \ref{tab:scoring}, demonstrate a notable superiority of our newly proposed baseline, LLM-Reviewer, across almost all evaluation aspects. This performance advantage is attributed to LLM-Reviewer's direct alignment with the objectives of the code review task, guided by specific quality criteria. In contrast, previous methods, steered by indirect text similarity metrics, often underperformed in several aspects, failing to achieve holistic excellence in comment quality.

\begin{table*}[ht]
\centering
\tiny
\caption{Average Scoring results by DeepCRCEval (both human and LLM evaluators).}
\label{tab:scoring}
\begin{tabular}{@{}lrrrrrrrrrrrrrrrrrr@{}}
\toprule
\multirow{3}{*}{Method Name} & \multicolumn{18}{c}{Scoring (1-10,   Higher = Better) (H. represents Humans and M. represents LLMs)}   \\ \cmidrule(l){2-19}   
                       & \multicolumn{2}{c}{C1} & \multicolumn{2}{c}{C2}  & \multicolumn{2}{c}{C3} & \multicolumn{2}{c}{C4} & \multicolumn{2}{c}{C5} & \multicolumn{2}{c}{C6} & \multicolumn{2}{c}{C7} & \multicolumn{2}{c}{C8} & \multicolumn{2}{c}{C9}   \\ \cmidrule(lr){2-3} \cmidrule(lr){4-5} \cmidrule(lr){6-7} \cmidrule(lr){8-9} \cmidrule(lr){10-11} \cmidrule(lr){12-13} \cmidrule(lr){14-15} \cmidrule(lr){16-17} \cmidrule(l){18-19}
                       & H.             & M.             & H.             & M.            & H.                                      & M.                                      & H.                                        & M.                                       & H.              & M.              & H.              & M.             & H.             & M.             & H.                                      & M.                                      & H.            & M.          \\ \midrule
Tuano           & 8.33           & 6.36           & 4.97           & 3.65          & 1.40                                    & 3.44                                    & 1.87                                      & 3.35                                     & 2.03            & 3.54            & 1.87            & 3.27           & 2.13           & 3.57           & 2.13                                    & 4.26                                    & 8.87          & 8.16         \\
CommentFinder          & 8.27           & 6.52           & 2.23           & 2.77          & 1.43                                    & 2.76                                    & 1.27                                      & 2.54                                     & 1.20            & 2.71            & 1.23            & 2.53           & 1.30           & 2.76           & 1.30                                    & 3.43                                    & 8.83          & 8.29         \\
CodeReviewer           & 7.83           & 7.13           & 5.93           & 3.68          & 1.00                                    & 3.38                                    & 2.27                                      & 3.29                                     & 1.77            & 3.53            & 1.90            & 3.17           & 2.50           & 3.54           & 3.20                                    & 4.35                                    & 9.50          & 8.37         \\
AUGER                  & 8.33           & 4.98           & 1.93           & 2.27          & 1.00                                    & 2.16                                    & 1.00                                      & 2.04                                     & 1.13            & 2.09            & 1.17            & 2.04           & 1.00           & 2.19           & 1.07                                    & 2.84                                    & 9.17          & 8.43         \\
CCT5                   & 9.63           & 7.53           & 3.30           & 2.90          & 1.00                                    & 2.66                                    & 1.23                                      & 2.55                                     & 1.03            & 2.60            & 1.00            & 2.51           & 1.23           & 2.77           & 1.70                                    & 3.79                                    & 9.83          & \textbf{9.27} \\ \midrule
\textbf{LLM-Reviewer}  & \textbf{9.97}  & \textbf{9.24}  & \textbf{10.00} & \textbf{9.55} & \textbf{9.67}                           & \textbf{9.17}                           & \textbf{9.80}                             & \textbf{9.40}                            & \textbf{9.83}   & \textbf{9.12}   & \textbf{9.37}   & \textbf{8.89}  & \textbf{9.87}  & \textbf{9.32}  & \textbf{9.90}                           & \textbf{9.59}                           & \textbf{9.97} & 8.23    \\ \bottomrule
\end{tabular}
% \vspace{-1cm}
\end{table*}

\begin{table}[htbp]
\centering
\scriptsize
\caption{Average ranking results by DeepCRCEval (both human and LLM evaluators).}
\label{table:ranking}
\begin{tabular}{@{}lrrrrrr@{}}
\toprule
      & \textbf{Tuano et al.} & \textbf{CommentFinder} & \textbf{CodeReviewer} & \textbf{AUGER} & \textbf{CCT5} & \textbf{LLM-Reviewer} \\ \midrule
Humans & 2.77         & 4.77          & 2.67         & 5.77  & 4.03 & \textbf{1}            \\
LLMs  & 3.3          & 4.17          & 3.33         & 5.19  & 4.00    & \textbf{1}            \\ \bottomrule
\end{tabular}
% \vspace{-1cm}
\end{table}

\subsubsection{Ranking}
The ranking results, depicted in the second part of Table \ref{table:ranking}, unequivocally place LLM-Reviewer at the top, as acknowledged by both human and LLM evaluators. Following LLM-Reviewer, Tufano and CodeReviewer were closely matched in quality, occupying the second and third positions, respectively. CCT5 was ranked fourth, CommentFinder fifth, and AUGER lagged at the sixth position.

\begin{tcolorbox}
    \textbf{Response to RQ3:} LLM-Reviewer, guided by direct task objectives and explicit criteria, excels in generating high-quality review comments, surpassing previous models.
\end{tcolorbox}

\section{Discussion}

\subsection{Implications}

Our findings highlight the limitations of existing text similarity metrics in evaluating code review comment generation. Our proposed framework, DeepCRCEval, demonstrates superior ability in discriminating between high and low-quality reviews, offering a more comprehensive assessment. Additionally, our new baseline, LLM-Reviewer, guided by direct targets and specific criteria, outperforms previous models.

\begin{figure}[t]
    \centering
    \includegraphics[width=0.7\linewidth]{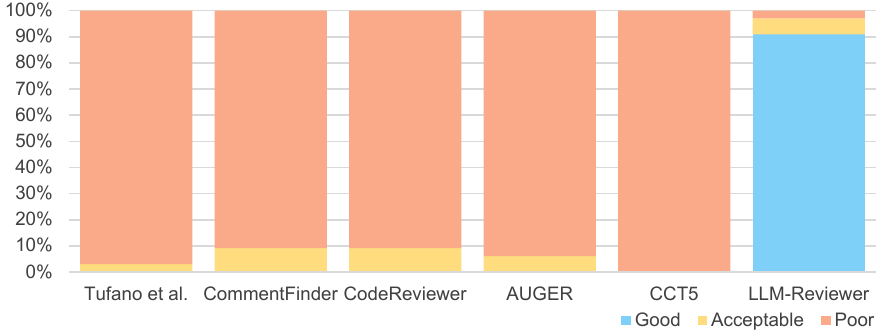}
    \caption{User feedback ratings distribution for ``Good'', ``Acceptable'', and ``Poor''.}
    \label{fig:user_feedback}
\end{figure}

To assess its practical utility, we tested a web application developed using the Gradio library with 5 industry developers. Users were instructed to input code snippets, and the application provided generated comments from each model. User feedback on total 66 cases, categorized as ``Good'' (I), ``Acceptable'' (II), or ``Poor'' (III), is summarized in Figure \ref{fig:user_feedback}. The feedback corroborates our findings: comments generated by LLM-Reviewer, noted for their quality and effectiveness, received more ``Good'' ratings. This user input highlights the potential direction for future research, emphasizing the need to delve into deeper, domain-specific features rather than treating code review as a standard text-to-text task.

\begin{tcolorbox}
    \textbf{Implications:} For code review comment generation, researchers should prioritize the fundamental objectives of code reviews instead of focusing merely on textual similarity. Incorporating the specific goals of code reviews directly into the training methodologies seems promising.
\end{tcolorbox}

\subsection{Case Study}

For a qualitative comparison, we present two case studies in Table \ref{table:case_study}—one highlighting a defect and another focusing on code improvement. These cases contrast high-quality comments from LLM-Reviewer with those of lower quality from other baselines. Comments from baseline models often suffer from a lack of relevance or generality, potentially confusing developers. This deficiency may stem from their reliance on indirect text similarity metrics, leading to suboptimal outcomes. Conversely, LLM-Reviewer's comments are notably superior, providing precise, clear explanations of problems, their causes, and suggested solutions, corroborating our previous findings.

\begin{lrbox}{\codeboxOne}
\begin{lstlisting}
public class SyncExample {
    private final Object lock1 = new Object();
    private final Object lock2 = new Object();
    public void method1() {
        synchronized(lock1) {
            synchronized(lock2) {
                System.out.println("Lock1 then Lock2");
            }
        }
    }
    public void method2() {
        synchronized(lock2) {
            synchronized(lock1) {
                System.out.println("Lock2 then Lock1");
            }
        }
    }
}
\end{lstlisting}
\end{lrbox}

\begin{lrbox}{\codeboxTwo}
\begin{lstlisting}
public void processMap(Map<String, Integer> map, String key) {
    if (map.containsKey(key)) {
        Integer value = map.get(key);
        System.out.println(value);
    }
}
\end{lstlisting}
\end{lrbox}

\begin{table}[ht]
\centering
\tiny
\caption{Case studies comparing baseline comments.}
\label{table:case_study}
\centering

% Case 1
\begin{tabular}{|p{2cm}|p{9.5cm}|}
\hline
\multicolumn{2}{|c|}{\textbf{Case 1: DeadLock Issue}} \\
\hline
\multicolumn{2}{|l|}{\textbf{Code:}} \\ \hline
\multicolumn{2}{|p{7.5cm}|}{
\resizebox{0.715\linewidth}{!}{\usebox{\codeboxOne}}
} \\
\hline
\textbf{Source} & \textbf{Comment} \\
\hline
Tufano et al. & This is exactly the same class, you can't have two methods, one with the same name, and another with the same name. \\
\hline
CommentFinder & May want to get rid of this System.out running loose here in this test case. \\
\hline
CodeReviewer & Why is this needed? \\
\hline
AUGER & yes i think we should have this in the hoosier state lock block \\
\hline
CCT5 & Remove this line. \\
\hline
\textbf{LLM-Reviewer} & \textbf{In ``SyncExample'' class, ``method1'' and ``method2'' obtain locks in a different order. If they are called concurrently, it can lead to a deadlock where each method is waiting for the other to release a lock. Always acquire locks in the same order to prevent deadlocks.}\\
\hline
\hline
\multicolumn{2}{|c|}{\textbf{Case 2: Map Performance Issue}} \\
\hline
\multicolumn{2}{|l|}{\textbf{Code:}} \\ \hline
\multicolumn{2}{|p{7.5cm}|}{
\resizebox{0.715\linewidth}{!}{\usebox{\codeboxTwo}}
} \\
\hline
\textbf{Source} & \textbf{Comment} \\
\hline
Tufano et al. & Can we make this a logger? \\
\hline
CommentFinder & > I would rather use the first version and avoid the cast to M. Is that OK? Please cast (M) map.put(key, value). We save an additional Tuple instance which scales for many calls. I see our collections as low-level, they should be as performant as possible. A cast is fast. \\
\hline
CodeReviewer & Remove this file, its irrelevant to this PR. \\
\hline
AUGER & why do we need this \\
\hline
CCT5 & Please remove this. \\
\hline
\textbf{LLM-Reviewer} & \textbf{The ``processMap'' method performs two lookups: one for ``containsKey'' and another for ``get''. This is inefficient. Instead, use ``get'' directly and check if the result is null. It performs the operation in a single lookup, improving efficiency.}\\
\hline
\end{tabular}
% \vspace{-0.3cm}
\end{table}

% \begin{figure}[ht]
%     \centering
%     \includegraphics[width=\linewidth]{images/case_study.pdf}
%     \caption{Case studies comparing baseline comments.}
%     \label{fig:case_study}
% \end{figure}

\begin{tcolorbox}
    \textbf{Case Study:} LLM-Reviewer's high-quality comments deliver concise, clear explanations of issues, their roots, and potential fixes, contrasting with the irrelevant or generic comments from other models.
\end{tcolorbox}

\subsection{Threats to Validity}

Several factors are crucial for assessing the validity of our findings. Firstly, our selection of large language models (LLMs), specifically GPT-4, was a deliberate decision. For DeepCRCEval, GPT-4 was chosen to emulate human evaluators because of its advanced capabilities. Likewise, GPT-4 served as the foundation for LLM-Reviewer due to its status as a leading and representative LLM. Another consideration is the focus on the Java programming language for our code review task. While Java has specific characteristics that differ from other languages, it is the most commonly used language in prior research, making it suitable for comparison. Finally, the scope of our reevaluation needs acknowledgment. We used graduate computer science students as proxies for actual developers in human evaluations. These students have significant programming experience, making them a reasonable approximation. Additionally, due to the high costs of manual analysis, our study covered a relatively small sample size. To mitigate this, we used LLM evaluators for broader analysis and compared the concordance between human and LLM evaluations, ensuring a thorough and robust assessment. 

However, using LLMs to evaluate LLMs still potentially introduces bias, which is why we also used human evaluators. The introduction of LLM-Reviewer aims to highlight the significant room for improvement in existing CRCGs. The results show that they score particularly low compared to LLM-Reviewer, a discrepancy even more pronounced when by humans, which cannot be wholly attributed to bias. 

\section{Related Work} \label{sec: related work}

The automation and evaluation of machine code review comments are recent developments, aligning with longstanding research interests in assessing the quality of human-written comments. Traditional evaluations predominantly focused on the ``usefulness'' of comments. Early methods, exemplified by Bosu et al. \cite{bosu2015characteristics}, employed decision trees and hand-crafted rules to categorize comments as ``useful'' or ``not useful'', often based on subsequent code modifications or ``wontfix'' labels. Rahman et al. \cite{rahman2017predicting} refined this approach by emphasizing comments' ``change-triggering'' characteristics, incorporating textual content and reviewer experience into their predictive models. Hasan et al. \cite{hasan2021using} expanded this further by integrating additional features from review contexts and reviewer backgrounds. A notable advancement came with Yang et al. \cite{yang2023evacrc}, who introduced a BERT-based scoring system across four dimensions (emotion, question, evaluation, and suggestion), marking a shift towards a more detailed and explanatory evaluation, beyond extensive feature engineering.

However, Yang et al.'s methodology diverges from ours in two key respects. First, their model assesses human-generated comments, whereas our focus is on \textbf{machine}-generated comments aimed at enhancing code review quality. This necessitates a more granular evaluation, emphasizing clarity and effectiveness in addressing actual defects or improvements, as detailed in Section \ref{sec: CRCGs}. Second, our approach demands a deeper evaluation, analyzing the interaction between \textbf{code and comment pairs}, while previous research primarily targeted comments alone. Without including code as a target, it is impossible to judge perspectives related to actual issues in the code. Finally, we utilize the emergent abilities of \textbf{LLMs} like in-context learning. Earlier models like BERT lacked the depth required for such analysis, but recent LLM advancements, especially GPT-4, have shown near-human comprehension in understanding both code and language. Hence, DeepCRCEval incorporates LLM evaluators to complement human evaluation, balancing reliability with reduced time and cost.

\section{Conclusion} \label{sec: conclusion}

This study challenges the prevailing evaluation methodology for CRCGs, arguing that text similarity metrics like BLEU and ROUGE-L are inadequate due to the questionable quality and validity of benchmark comments. As a solution, we introduced DeepCRCEval, a framework directly addressing developers' concerns, and LLM-Reviewer, a lightweight, training-free baseline for CRCG evaluation. LLM-Reviewer, guided by clear and direct task goals and criteria, contrasts with methods relying on text similarity for training. 

Our empirical findings suggest that CRCGs might overstate their improvements when focused on text similarity metrics, often producing comments that are irrelevant or overly generic. In contrast, LLM-Reviewer demonstrates the ability to provide clear, concise explanations of issues, their causes, and potential solutions, even without specific training. DeepCRCEval offers superior discrimination and comprehensiveness compared to previous metrics, and significantly reduces costs while maintaining reliable evaluation standards when employing LLMs as alternative evaluators. Our work lays a foundation for task-specific evaluations for code review comment generation, and highlights that future researchers should not neglect the original objectives of the code review.

\bibliographystyle{splncs04}
\bibliography{reference}

\appendix
\section{Additional Background} \label{sec: additional background}
\subsection{Modern Code Review}

Modern Code Review (MCR) has become an integral part of software development. As illustrated in Figure \ref{fig:code_review}, this process primarily comprises two elements:

\begin{figure}[htbp]
    \centering
    \includegraphics[width=0.8\linewidth]{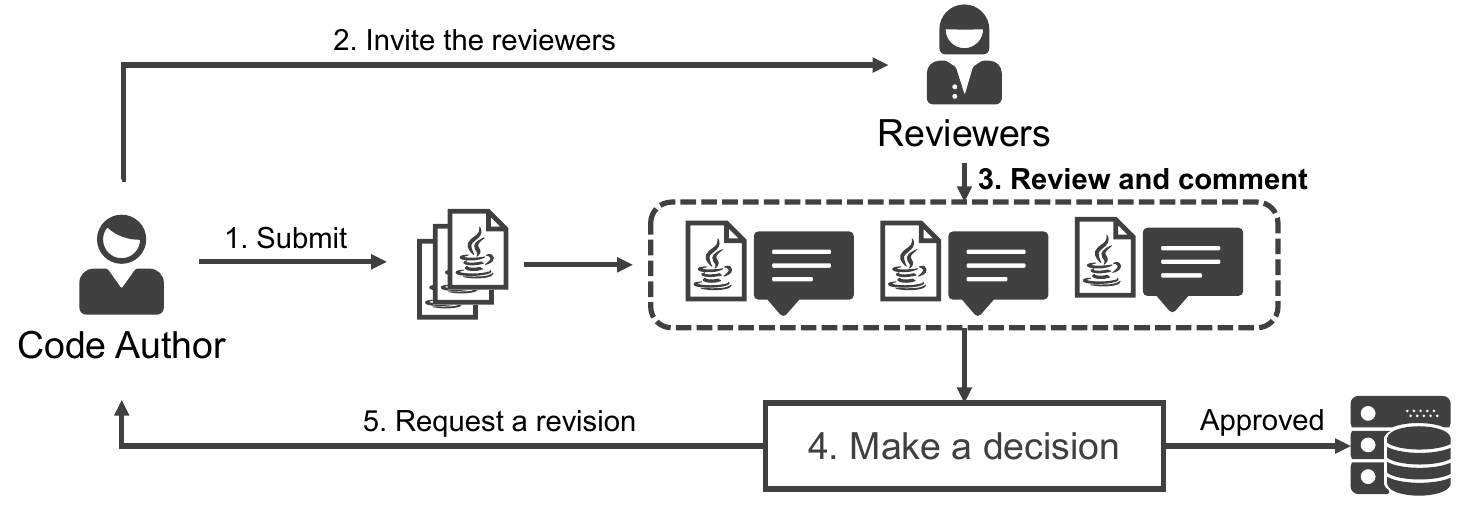}
    \caption{The overall workflow of the code review process.}
    \label{fig:code_review}
\end{figure}

\begin{enumerate}[label=$\bullet$, leftmargin=1em, labelsep=0.5em]
\item \textbf{Submitted Code Snippet.} The committer submits new code for review. This step is crucial as it introduces changes to the existing codebase, necessitating thorough examination.
\item \textbf{Review Comment.} These are insights provided by the reviewers. They not only pinpoint issues in the submitted code but also guide the committer towards understanding the underlying problems and potential fixes.
\end{enumerate}

\subsection{Task Description of Code Review Comment Generation} \label{sec: code review comment generation}

The goal of automated code review comment generation is to either augment or replace human effort in the code review process, thereby reducing labor costs. The task involves generating a pertinent comment \( y \) for a given code snippet \( x \). This comment should effectively and succinctly highlight any issues present. During training, the model learns to estimate the probability \( P(y | x) = \prod_{i=1}^n P(y_i | y_{<i}, x) \), where \( y_i \) represents the \( i \)-th token in the comment, and \( y_{<i} \) encompasses all preceding tokens. In the inference phase, the model generates a comment \( y' \) based on the probability \( P(y' | x) \).

\section{Finding Quality Evaluation Criteria} \label{sec: finding quality evaluation criteria}

\begin{figure}[htbp]
    \centering
    \includegraphics[width=0.8\linewidth]{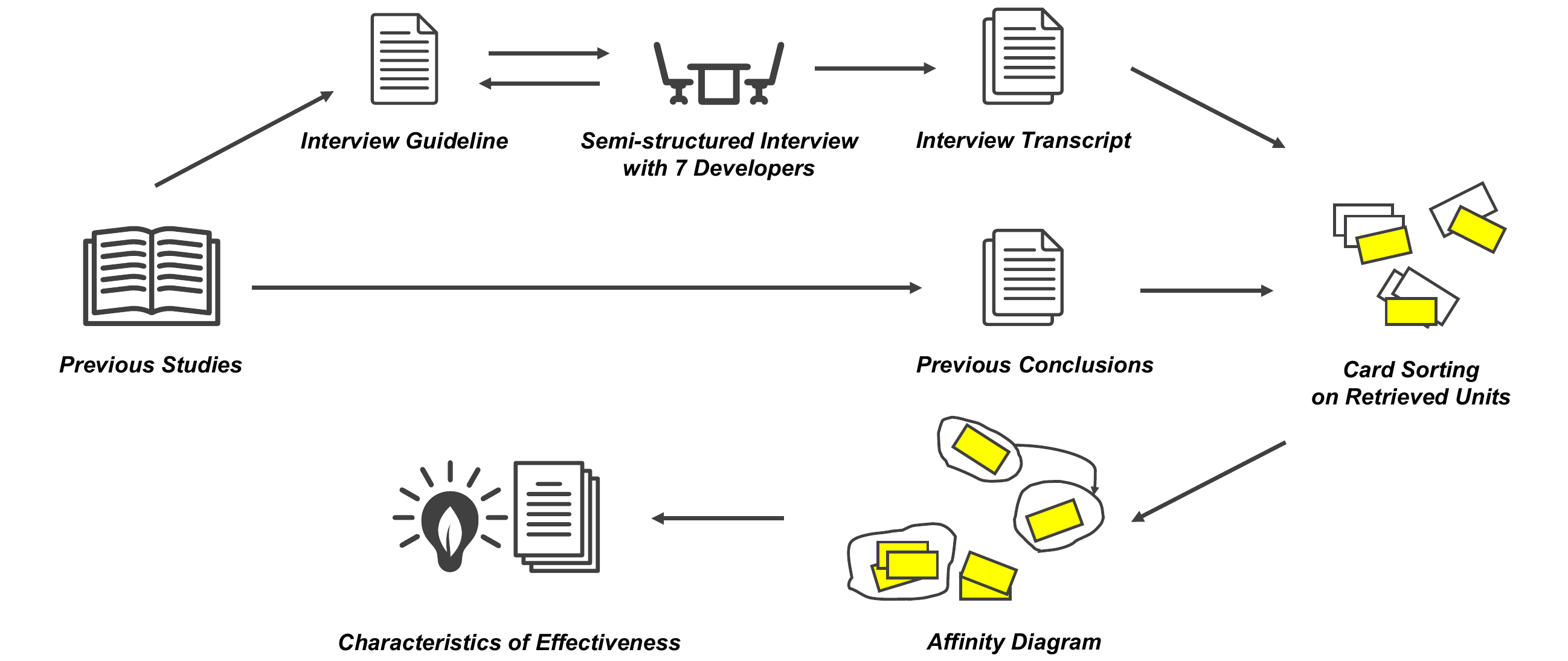}
    \caption{Process of defining quality standards for comments.}
    \label{fig:characteristics}
\end{figure}

Before analyzing the benchmark comments, we first need to establish the characteristics of a high-quality comment. Our approach to defining effective code review comments combines qualitative and quantitative methods \cite{creswell2017research}, drawing on varied data sources for comprehensive insights (Figure \ref{fig:characteristics} depicts this methodology), with the results presented in Section \ref{sec: quality}, which is consistent with the previous research \cite{kononenko2016code}.

\subsection{Review of Previous Studies}

Initially, we reviewed existing research by Kononenko et al. \cite{kononenko2016code}, which delineates developers' perspectives on high-quality code review comments. This review highlighted attributes such as clarity, relevance beyond mere code styling, constructive feedback, reviewer expertise, and mentoring potential. These elements informed the creation of our interview guidelines, emphasizing aspects like readability, relevance, problem identification, completeness, actionability, specificity, and clarity in explanations.

\subsection{Semi-structured Interviews with Developers}

Based on the former results, we conducted extended semi-structured interviews \cite{lindlof2017qualitative} with seven industry developers, each with over five years of experience and familiarity with machine learning tools in software engineering. These interviews, each lasting 10-15 minutes, allowed for iterative refinement of our guidelines. Saturation in responses was observed after 3-4 interviews, suggesting a consistency in the insights provided.

\subsection{Card Sort and Affinity Diagram}

Subsequently, we applied an open card sort technique to categorize interview data, followed by the use of affinity diagrams to connect related concepts. This process, driven by consensus among the authors, led to the emergence of two additional aspects: contextual adequacy and brevity, underlining the importance of locating problems quickly and favoring concise comments.

\section{Detailed Analysis Methodology for Dateset Comment Quality} \label{sec: Detailed Analysis Methodology for Dateset Comment Quality}

\subsection{Quality}

To assess the quality of the reference comments, we implemented a human scoring system. Insights from our semi-structured interviews informed the development of an efficient scoring process. Interviewees recommended features for a scoring tool that enhances usability, such as pause-and-resume functionality, resilience to network disruptions, and clear guidelines. Consequently, we developed an offline scoring application (an executable file created using QT), embedding these suggested features along with a cumulative timing function. This tool is accessible in our open-source repository. Five master's and doctoral students in computer science, each with over six years of programming experience, conducted the scoring. Each comment was evaluated on a 1-10 scale, where higher scores denote better quality.

\subsection{Category}

To analyze the distribution of comment categories in OSS projects, we adapted the Delphi Method \cite{dalkey1963experimental}. This categorization was carried out by five computer science master's/doctoral students, all seasoned programmers with at least six years of experience. The process involved:
\begin{enumerate}
    \item \textbf{Individual Assessment:} Participants independently categorized each comment into one of nine predefined categories, ensuring a breadth of perspectives.
    \item \textbf{Group Deliberation:} The team then convened to discuss and resolve any differences, especially in instances where initial classifications lacked majority agreement.
    \item \textbf{Consensus Building:} Through iterative discussions and re-assessments, the group aimed to achieve a consensus on the categorization of each comment.
\end{enumerate}

\subsection{Tone and Context}

The evaluation of tone and context was conducted using the aforementioned Nominal Group Technique (NGT) sessions, involving three authors of this study. This approach ensured a structured and collaborative analysis of these aspects.

\section{Detailed Prompt Templates}

\subsection{Prompt Template of LLM Evaluators}

The comprehensive prompt template utilized for LLM evaluators in our CRCEval framework is detailed in Table \ref{tab:prompt_template_eval}. This template is meticulously designed to guide LLMs effectively, integrating task-specific instructions, guidelines, and a generation format conducive to chain-of-thought (CoT) reasoning. This approach not only instructs the LLMs but also amplifies their analytical capabilities.

To ensure accuracy in the evaluation outcomes, the ranking results are regularized, addressing and correcting any minor formatting inconsistencies that may arise. As a similar variant, for the evaluation of benchmark comments, particularly in the context of one-to-one code-comment pairs, the ranking component of the prompt is excluded. Additionally, slight modifications are made to the scoring task description to tailor it to the specific needs of benchmark comment evaluation.

\begin{table}[htbp]
\tiny
\centering
\caption{Detailed prompt template for LLM evaluators.}
\label{tab:prompt_template_eval}
\begin{tabularx}{0.95\linewidth}{@{} >{\raggedright\arraybackslash}p{1.6cm} @{} X @{}}
\toprule
Element                     & Content \\ \midrule
Task Description (Scoring) & As a thorough and unbiased AI evaluator, you're tasked with ranking several AI models based on the quality of their code review comments for a given problematic Java code snippet. For each model, you will assign a score from 1-10 (higher scores indicate better performance) for each of the following nine metrics. Upon scoring, you will create an overall leaderboard for the models. \\ \midrule
Guidelines                  & Please note:\newline
- Avoid any positional bias. The order of comments must not influence ranking.\newline 
- The length of the comments should not affect your evaluation.\newline 
- Models' names should not influence your judgment.\newline 
- Maintain objectivity throughout the process.\newline\newline
The nine metrics are:\newline 
\{Nine Criteria in Section \ref{sec: quality}\} \\ \midrule
Evaluation Objects & Problematic Java code snippet: \newline \{code\}\newline 
Comments from the models: \newline \{comment\_list\} \\ \midrule
Task Description (Ranking) & After scoring, rank the models by their overall performance quality. A rank of 1 signifies the best output. If models tie, assign them the average rank corresponding to their position. For example, if two models tie for first place, both receive a rank of 1.5, and the next model gets a rank of 3. If three models tie for second place, all are ranked 3, and the next model, if any, is ranked 5. \\ \midrule
Generation Format & Structure your output as follows:\newline
\#\#\# Scoring: \newline 
[
    \{"model": \textless{}model-name\textgreater{}, "score": [list of scores in order]\}\newline
    \{"model": \textless{}model-name\textgreater{}, "score": [list of scores in order]\}, 
    ...
]\newline\newline
\#\#\# Chain-of-Thoughts:\newline
Provide a short explanation for your ranking\newline\newline 
\#\#\# Ranking: \newline 
[ 
    \{"model": \textless{}model-name\textgreater{}, "rank": \textless{}model-rank\textgreater{}\}\newline
    \{"model": \textless{}model-name\textgreater{}, "rank": \textless{}model-rank\textgreater{}\}, 
    ...
]\newline\newline
The sections "Scoring" and "Ranking" must be valid Python dictionary lists, ready to be directly executed in Python. Each section should begin with its respective title, exclusively: "\#\#\# Scoring:", "\#\#\# Chain-of-Thoughts:", and "\#\#\# Ranking:". The goal is to produce a ranking human evaluators would agree with. \\
\bottomrule
\end{tabularx}
\end{table}

\subsection{Prompt Template of LLM-Reviewer}

In deploying LLM-Reviewer, we harness the few-shot learning capabilities of LLMs through a meticulously crafted in-context prompt template. This template, detailed in Table \ref{tab:prompt_template_llm_reviewer}, is structured into three main components to optimize the LLM's performance:

\begin{table}[htbp]
\tiny
\centering
\caption{Detailed prompt template for LLM-Reviewer.}
\label{tab:prompt_template_llm_reviewer}
\begin{tabularx}{0.95\linewidth}{@{} >{\raggedright\arraybackslash}p{1.6cm} @{} X @{}}
\toprule
Element & Content \\ \midrule
Task Description & As a thorough AI code reviewer, you're tasked to review several Java code snippets. Each code snippet may contain one or more issues. For each code snippet, you should provide a succinct explanatory comment according to the following guidelines. \\ \midrule
Guidelines & Guidelines:\newline
The comment should adhere to the following criteria:\newline
\{Nine Criteria in Section \ref{sec: quality}\} \\ \midrule
K Demonstrations & \#\#\#\newline
1. Java code snippet: \newline\{Code snippet of Demonstration 1\}\newline
1. Comment: \newline\{Comment of Demonstration 1\}\newline
\#\#\#\newline
\{Demonstration 2 \& 3\}\newline
\#\#\#\newline
4. Java code snippet: \newline\{Target Code snippet\}\newline
4. Comment: \\
\bottomrule
\end{tabularx}
\end{table}

\begin{enumerate}[label=\arabic*., labelwidth=*, labelsep=0em, leftmargin=!, align=left, start=1]
    \item Task Description: This section provides a clear, concise directive for the LLM, outlining the specific task of generating a code review comment. It sets the context and objective for the LLM, ensuring its outputs are aligned with the desired outcomes.
    \item Guidelines: These are carefully formulated instructions that include notes on the expected format and content of the review comments. The guidelines serve to steer the LLM towards generating relevant, useful, and context-appropriate comments.
    \item Exemplar Demonstrations: To leverage the LLM's few-shot learning ability, the template includes a set of exemplar demonstrations. These are carefully selected examples that illustrate the kind of output desired from the LLM, serving as a reference point for its comment generation process.
\end{enumerate}

This prompt is then integrated with the code snippet under review before being fed into the LLM for comment generation.

\section{Benchmark Comment Quality Analysis Results by LLM Evaluators}

For the analysis of benchmark comments in Section \ref{sec: analysis}, in addition to human scoring, our study also utilized the LLM evaluators introduced in Section \ref{sec: llm evaluators} for a broader analysis. We expanded our evaluation to encompass 1,000 cases per dataset, with the findings by LLM evaluators detailed in Table \ref{tab:quality_by_llms}. While there are variances in specific values, the overall trends observed by LLM evaluators align with those identified by human evaluators. Generally, LLM evaluators exhibited a greater tolerance across most criteria but concurred with human evaluators that benchmark comments largely fall short in criteria C2-C8. A notable exception was observed in the aspect of completeness, where LLM evaluators showed significantly different results. This divergence is attributed to the challenge human evaluators face in identifying unmentioned issues within a limited timeframe, highlighting one of the key reasons for automating code review comment generation.

\begin{table}[t]
\centering
\tiny
\caption{Average quality of comments (\ding{182} upper part, 1-10) and percentage of low-quality cases (\ding{183} lower part, 0\%-100\%) in OSS datasets by LLM evaluators. C1-C9 represent criteria mentioned in Section \ref{sec: quality}.}
\label{tab:quality_by_llms}
\begin{tabular}{@{}llrrrrrrrrr@{}}
\toprule
                    & Dataset & C1   & {\color[HTML]{FF0000}C2}   & {\color[HTML]{FF0000} {C3}}   & {\color[HTML]{FF0000} {C4}}   & {\color[HTML]{FF0000} {C5}}   & {\color[HTML]{FF0000}C6}   & {\color[HTML]{FF0000} {C7}}   & {\color[HTML]{FF0000} {C8}}   & C9   \\ \midrule
                    & Tufano  & 8.05 & {\color[HTML]{FF0000}6.72} & {\color[HTML]{FF0000} {5.23}} & {\color[HTML]{FF0000} {5.59}} & {\color[HTML]{FF0000} {5.85}} & {\color[HTML]{FF0000}4.56} & {\color[HTML]{FF0000} {6.03}} & {\color[HTML]{FF0000} {6.02}} & 9.39 \\
\multirow{-2}{*}{\ding{182}} & Crer    & 8.75 & {\color[HTML]{FF0000}7.91} & {\color[HTML]{FF0000} {6.17}} & {\color[HTML]{FF0000} {6.51}} & {\color[HTML]{FF0000} {6.38}} & {\color[HTML]{FF0000}5.74} & {\color[HTML]{FF0000} {7.18}} & {\color[HTML]{FF0000} {7.14}} & 9.58 \\ \midrule
                    & Tufano  & 3.6\%  & {\color[HTML]{FF0000}28.6\%} & {\color[HTML]{FF0000} {48.4\%}} & {\color[HTML]{FF0000} {42.0\%}} & {\color[HTML]{FF0000} {41.4\%}} & {\color[HTML]{FF0000}61.0\%} & {\color[HTML]{FF0000} {36.6\%}} & {\color[HTML]{FF0000} {37.0\%}} & 1.0\%  \\
\multirow{-2}{*}{\ding{183}} & Crer    & 0.8\%  & {\color[HTML]{FF0000}12.4\%} & {\color[HTML]{FF0000} {33.1\%}} & {\color[HTML]{FF0000} {30.7\%}} & {\color[HTML]{FF0000} {33.5\%}} & {\color[HTML]{FF0000}39.5\%} & {\color[HTML]{FF0000} {22.9\%}} & {\color[HTML]{FF0000} {20.4\%}} & 0.3\%  \\ \bottomrule
\end{tabular}
\end{table}

\end{document}